\documentclass[11pt]{article}

\usepackage[final]{acl}

\usepackage{times}
\usepackage{latexsym}

\usepackage[T1]{fontenc}

\usepackage[utf8]{inputenc}

\usepackage{microtype}

\usepackage{inconsolata}

\usepackage{graphicx}
\usepackage{amsmath,amssymb,amsthm}
\usepackage{algorithm}
\usepackage{algorithmic}
\usepackage{booktabs}
\usepackage{multirow}
\usepackage[table]{xcolor}
\usepackage{makecell}
\usepackage{authblk}
\title{SSG: Logit-Balanced Vocabulary Partitioning for LLM Watermarking}

\author{\textbf{Chenxi Gu}, \textbf{Xiaoning Du}\thanks{Corresponding author.}, \textbf{John Grundy}}
\affil{Monash University}
\affil{\texttt{\{chenxi.gu, xiaoning.du, john.grundy\}@monash.edu}}

\begin{document}
\maketitle
\begin{abstract}
Watermarking has emerged as a promising technique for tracing the authorship of content generated by large language models (LLMs). Among existing approaches, the KGW scheme is particularly attractive due to its versatility, efficiency, and effectiveness in natural language generation.
However, KGW's effectiveness degrades significantly under low-entropy settings such as code generation and mathematical reasoning. A crucial step in the KGW method is random vocabulary partitioning, which enables adjustments to token selection based on specific preferences. Our study revealed that the next-token probability distribution plays an critical role in determining how much, or even whether, we can modify token selection and, consequently, the effectiveness of watermarking.
We refer to this characteristic, associated with the probability distribution of each token prediction, as \emph{watermark strength.} In cases of random vocabulary partitioning, the lower bound of watermark strength is dictated by the next-token probability distribution. However, we found that, by redesigning the vocabulary partitioning algorithm, we can potentially raise this lower bound. In this paper, we propose SSG (\textbf{S}ort-then-\textbf{S}plit by \textbf{G}roups), a method that partitions the vocabulary into two logit-balanced subsets. This design lifts the lower bound of watermark strength for each token prediction, thereby improving watermark detectability. Experiments on code generation and mathematical reasoning datasets demonstrate the effectiveness of SSG. The source code is available at \url{https://github.com/AllenG-L/SSG}.


\end{abstract}

\section{Introduction}
Large language models (LLMs) can generate high-quality, human-like content, making them powerful tools for diverse generative tasks. Yet these capabilities also pose risks: LLM outputs may be misused to infringe on proprietary interests, circumvent usage policies, or spread harmful misinformation. For example, most proprietary providers such as OpenAI, Anthropic, and Google Gemini prohibit the use of their outputs for training other models \cite{openai,gemini,anthropic}, but proving such violations is difficult when generated text cannot be reliably distinguished from human-written or non-watermarked model outputs. Similarly, malicious actors may employ LLMs to generate convincing fake news \cite{liu2024surveytextwatermarkingera}, urging regulatory bodies in the US, EU, China, and G7 countries to seek robust mechanisms for content authentication and provenance \cite{eu}. These challenges highlight the urgent need for reliable watermarking techniques.  

Among existing approaches, the KGW watermarking scheme has attracted attention for its simplicity, efficiency, and effectiveness in natural language generation \cite{kirchenbauer2024watermarklargelanguagemodels,liu2024surveytextwatermarkingera}. KGW partitions the vocabulary into two subsets (green and red) based on a context-dependent hash and biases the green subset's logits to embed a hidden statistical signal. A standard statistical test, such as a z-score test, can then detect this signal by measuring the overrepresentation of green tokens.  

However, KGW watermarking degrades in low-entropy settings such as code generation \cite{lee2024wrotecodewatermarkingcode,lu2024entropybasedtextwatermarkingdetection,huang-etal-2025-low}, where the next-token probability distribution is sharply peaked at a few tokens. 
While investigating this limitation, we found that next token probability distribution inherently affect how much, or even whether, we can effectively bias the green set of tokens, and consequently, the watermarking effectiveness.
We refer to this characteristic associated with the probability distribution of next token prediction as \emph{watermark strength.}
Formally, it quantifies the normalized probability shift toward the green set at next token prediction.
Under random vocabulary partitioning - the strategy hired by KGW, the lower bound of watermark strength is dictated by the next-token probability distribution (detailed analysis provided in Section~\ref{motivation}).
Low watermark strength is usually observed when most high-logit tokens are in the same subset—a situation that arises non-negligibly in low-entropy tasks where the number of high-logit tokens is quite limited.



To mitigate this problem, we propose \textbf{SSG (Sort-then-Split by Groups)}, a novel algorithm for improving the lower bound of \emph{watermark strength,} thereby enhancing watermark detectability.  
SSG sorts tokens by their logits, groups them, and then applies a hash-based split such that the green and red sets are approximately logit-balanced. 
By doing so, SSG guarantees a higher lower bound on watermark strength than KGW, and the proof is presented in Section~\ref{AnalysisSSG}. Moreover, this proof extends to high-entropy tasks, demonstrating SSG's generalizability across all generation scenarios. Experiments on multiple datasets demonstrate that SSG can enhance watermark detectability on different tasks without compromising text quality.

Our main contributions are as follows:  
\begin{itemize}
    \item We analyze the limitations of KGW watermarking in low-entropy settings and show why random vocabulary partitioning leads to weak watermark injection.  
    \item We introduce watermark strength, a token-level metric that provides fine-grained insight into watermark injection. 
    \item We propose \textbf{SSG}, a logit-balanced partitioning scheme that improves watermark detectability.  
    \item We conduct comprehensive evaluations on low-entropy and high-entropy tasks, demonstrating that SSG achieves superior detectability over baselines without degrading text quality.  
\end{itemize}

\section{Related Works}

Large language models (LLMs) such as GPT \cite{Radford2018ImprovingLU}, LLaMA \cite{touvron2023llamaopenefficientfoundation}, and GPT-4 \cite{openai2024gpt4technicalreport} have advanced rapidly in recent years, achieving state-of-the-art performance across diverse tasks, including machine translation \cite{openai2024gpt4technicalreport}, code generation \cite{li2023starcodersourceyou}, and many others \cite{touvron2023llamaopenefficientfoundation}. Alongside these successes, however, their widespread adoption raises concerns over potential misuse and intellectual property protection, like unauthorized dataset usage\cite{Sun_2023}. To address these challenges, watermarking has emerged as a promising approach for attributing and detecting LLM-generated content.

Watermarking methods can broadly be categorized into three classes: logit-based \cite{kirchenbauer2024watermarklargelanguagemodels,lee2024wrotecodewatermarkingcode,lu2024entropybasedtextwatermarkingdetection,chen2024watmelosslesswatermarkinglexical}, sampling-based \cite{christ2023undetectablewatermarkslanguagemodels,kuditipudi2024robustdistortionfreewatermarkslanguage,hou2024semstampsemanticwatermarkparaphrastic}, and training-based \cite{sun2022coprotectorprotectopensourcecode,Sun_2023,gu2024learnabilitywatermarkslanguagemodels}. Among these, logit-based approaches are particularly attractive due to their simplicity, versatility, and low computational cost \cite{liu2024surveytextwatermarkingera}.

A representative example is KGW \cite{kirchenbauer2024watermarklargelanguagemodels}, which partitions the vocabulary into “green” and “red” sets using a hash of the preceding context, and then adds a bias to the logits of green tokens. This encourages the model to generate more green tokens, enabling detection through a z-score statistical test. 

While effective for high-entropy text, KGW performance degrades substantially in low-entropy settings such as code generation, where the next-token distribution is concentrated on a small set of candidates \cite{lee2024wrotecodewatermarkingcode,lu2024entropybasedtextwatermarkingdetection}. In such cases, perturbing logits produces only a negligible distributional shift, weakening the watermark signal and reducing detection accuracy.
To mitigate this, \citet{lee2024wrotecodewatermarkingcode} propose restricting watermark injection to high-entropy tokens, while \citet{lu2024entropybasedtextwatermarkingdetection} introduce entropy-weighted scoring to prioritize high-entropy tokens during detection. 

Overall, these studies ~\cite{lee2024wrotecodewatermarkingcode,lu2024entropybasedtextwatermarkingdetection,huang-etal-2025-low}focus on improving the watermarking for low-entropy tasks via modifying the detection algorithm, while SSG aims to improve the injection algorithm.

A recent concurrent work \citet{Park_Park_Ahn_Han_2026} also proposes a design of logit-balanced vocabulary partitioning for effective LLM watermarking. 
Similarly, it sorts and groups all tokens by logits.
Unlike our strategy of assigning green and red colors independently for each group, \citet{Park_Park_Ahn_Han_2026} assigns colors uniformly across all groups.
Both approaches offer similar enhancements, but we also provide a thorough theoretical analysis of the performance improvements. 


\section{Preliminaries}

\textbf{Language Model:} A language model $\mathcal{M}$ is fundamentally a function for prediction the next token in a sequence. $\mathcal{M}$ takes an input sequence $\boldsymbol{x}=[x_0,...,x_{n-1}]$ and produces the logit $L_n$, which will be transformed into a probability distribution $P_n$ over a vocabulary set $\mathcal{V}$. Formally, we have $P_n=softmax(L_n)=softmax(\mathcal{M}(\boldsymbol{x_{0:n-1}}))$. The next token $y_n$ is selected based on $P_n$ by sampling. This generation process is auto-regressive until a whole sequence $\boldsymbol{y}$ is finished.

\textbf{Watermarking Algorithm} aims to embed hidden pattern into LLMs' output to help people identify its authorship. It comprises two stages: {Watermark Injection} and {Watermark Detection}. During {Watermark Injection}, the probability distribution output will be altered from $P_n$ to $\tilde{P}_n$, which is determined by a secret key $s$. As a result, the output sequence is changed to $\boldsymbol{\tilde{y}}$ instead of $\boldsymbol{y}$. At {Watermark Detection} stage, the detection algorithm takes a suspected sequence $\boldsymbol{x}^*$ as input, and determines if it is generated by $\mathcal{M}$.

\textbf{KGW scheme Watermarking Algorithm:} To embed watermark when predict $y_n$, KGW scheme watermarking methods first divide vocabulary $\mathcal{V}$ into green token set $\mathcal{V}^{g}$ with size of $\gamma*\mathcal{|V|}$ where $\gamma$ is green set ratio, and red token set $\mathcal{V}^{r}$, via a hash function determined by secret key $s$ and context $\boldsymbol{x}_{-h:}$ Then, the logits of token $v_i$ in $\mathcal{V}^{g}$, denoted as $L^{g}_{n,i}$, will be increased by logit bias term $\delta$, to get watermarked logit $\tilde{L}_n=[L^{g}_n+\delta,L^{r}_n]$. At last, next token $\tilde{y}_n$ will be sampled according to $\tilde{P}_n$. During watermark detection stage, for a suspected text $\boldsymbol{x}$, a statistical test is performed. The null hypothesis $\mathcal{H}_0$ is :\texttt{The text is not watermarked}. If the z-score of statistical test is above a predefined threshold $\tau$, $\mathcal{H}_0$ could be rejected and one can conclude that $\boldsymbol{x}$ is watermarked.

\textbf{Low-entropy Generation Tasks:} Low-entropy generation tasks refer to scenarios in which the language model’s next-token distribution is highly concentrated on only a few candidates. Formally, if the next-token distribution is denoted as $P = \{p_i\}$, the entropy is given by $H(P) = - \sum_i p_i \log p_i$. A low-entropy setting arises when $H(P)$ is small, meaning that most of the probability is assigned to a small subset of tokens, while the rest receive negligible likelihood. Typical examples include code generation, mathematical reasoning, structured formats generation(e.g., JSON, SQL, or HTML).

\section{Method}
This section provides a detailed description of SSG. In Section 4.1, we first explain why random vocabulary partition, which is adopted in prior KGW-scheme, induces the ineffectiveness in low-entropy settings. Then the whole pipeline of SSG watermark injection stage is described in Section 4.2. Section 4.3 details the process of watermark detection. Finally, we analyze SSG in Section 4.4 to show why it could achieve more reliable watermark injection.

\subsection{Motivation}
\label{motivation}
We begin by formalizing the Watermarking Strength, which is a token-level metric that highly-related to watermark detectability of KGW-scheme and then explain why KGW-scheme could have poor Watermarking Strength on low-entropy tasks. 
Let $L_i$ be the model logit 
for token $v_i \in \mathcal{V}$. The corresponding unnormalized weight after softmax is
\begin{equation}
w_i = e^{L_i}.
\end{equation}

Given a partition of $\mathcal{V}$ into a green set $\mathcal{V}^{g}$ and a red set $\mathcal{V}^{r}$, 
the total probability assigned to the green set before watermarking is
\begin{equation}
p_g = \sum_{v_i \in \mathcal{V}^{g}} \frac{w_i}{\sum_{v_j \in \mathcal{V}} w_j}.
\end{equation}

In the KGW scheme, a fixed bias $\delta > 0$ is added to the logits 
of all green tokens, yielding modified logits
\begin{equation}
\tilde{L}_i =
\begin{cases}
L_i + \delta, & v_i \in \mathcal{V}^{g}, \\
L_i, & v_i \in \mathcal{V}^{r}.
\end{cases}
\end{equation}

The total probability assigned to the green set after watermarking becomes
\begin{equation}
\tilde{p}_g = \frac{e^{\delta} p_g}{e^{\delta} p_g + (1 - p_g)}.
\end{equation}

We define the \emph{Watermarking Strength} as the normalized increase in the green 
total probability due to watermarking:
\begin{equation}
f_{ws}(\delta,p_g) = \frac{\tilde{p}_g - p_g}{\sqrt{p_g(1-p_g)}}.
\label{eq:strength}
\end{equation}

which could be simplified as:

\begin{equation}
f_{ws}(\delta,p_g)
= \frac{(e^{\delta} - 1)\,\sqrt{p_g(1-p_g)}}{1 + (e^{\delta} - 1)p_g}.
\label{eq:strength2}
\end{equation}

Intuitively, $f_{ws}$ measures how much the bias $\delta$ 
alters the model’s next-token distribution toward the green set. Thus, $f_{ws}$, which is a concave function about $\delta$ and $p_g$ depicted in Appendix \ref{gws} , serves as a quantitative measure of the effectiveness 
of watermark injection at each next-token generation step. The lower bound of it implies that the green set contains either the top or bottom 50\% of logit tokens. 

In low-entropy settings, suppose the top $m$ tokens have logits $L_{1}\ge\cdots\ge L_{m}\gg \sum_{i>m}L_i$.
With probability $2^{1-m}$, random vocabulary split places all $m$ top tokens into the \emph{same} subset, which happens non-negligibly when $m$ is small in low-entropy settings, and hence $p_g$ is close to $0$ or $1$ as $\sum_{i>m}p_i\to 0$. As a result, $f_{ws}$ could be nearly $0$ in Equation \ref{eq:strength2}, causing a ineffective watermark injection. 

To improve on $f_{ws}$, one option is increase the bias term $\delta$, however, this could induce a significant text quality drop \cite{kirchenbauer2024watermarklargelanguagemodels}. Instead, SSG aims to avoid such phenomenon by guaranteeing the tokens can be divided into two logit-balanced sets. By doing so, SSG enhances watermark detectability without comprising text quality.

\subsection{Watermark Injection}

Algorithm \ref{alg:ssg_partition} describes how SSG generate watermarked text with a language model $\mathcal{M}$. First, the model computes the logits $L=\mathcal{M}(\boldsymbol{x})$ for all vocabulary tokens. The tokens are then sorted in descending order of their logits, and the resulting sequence is partitioned into consecutive groups of size two. 

Each group is processed independently: using a pseudorandom generator seeded by the secret key $s$ combined with the token ids in every group $\mathcal{B}$, one token from the group is assigned to the green set $\mathcal{V}^g$, and the other to the red set $\mathcal{V}^r$. 
After partitioning, the logits of all tokens in the green set are increased by $\delta$, while the red tokens remain unchanged.
The modified logits $\tilde{L}$ are then passed through the softmax function to produce the biased probability distribution $\tilde{P}$, from which the next token $y$ is sampled.

Despite the strong effectiveness of SSG demonstrated in our experiments, the procedure of SSG on the whole vocabulary can be computationally expensive, as it requires sorting the logits at every token prediction step. To strike a balance between effectiveness and efficiency, we propose to conduct SSG only on top-$k$ logits tokens. The key observation is that in low-entropy settings, only a few tokens dominate the probability. Thus, ensuring that these high-logit tokens are evenly distributed between the green and red sets is sufficient to achieve a balanced vocabulary split. Based on this insight, SSG applies only on the top-$k$ highest-probability tokens, while randomly splitting the remaining tokens. Our experiments show that this approach substantially reduces computational cost while maintaining watermarking effectiveness.

\begin{algorithm}[t]
\caption{Single Step Watermark Injection of SSG}
\label{alg:ssg_partition}
\begin{algorithmic}[1]
\REQUIRE LLM $\mathcal{M}$, context $\boldsymbol{x}$, vocabulary $\mathcal{V}$, secret key $s$, watermark window size $h$, logit bias $\delta$, Green list ratio $\gamma$, Top-k hyperparameter $k$
\ENSURE Next token $y$, Green set $\mathcal{V}^g$, Red Set $\mathcal{V}^r$
\STATE $L \gets \mathcal{M}(\boldsymbol{x})$ \hfill 
\STATE $\pi \gets$ indices of $\mathcal{V}$ sorted by $L_i$ in descending order
\STATE Partition $\pi_{0:k-1}$ into consecutive groups $\mathcal{B}_1,\mathcal{B}_2,\dots,\mathcal{B}_{k/2}$ of size 2
\STATE $\mathcal{V}^g \gets \emptyset$, $\mathcal{V}^r \gets \emptyset$
\FOR{each group $\mathcal{B}$}
    \STATE $(v_1, v_2) \gets$ tokens in $\mathcal{B}$
    \STATE Seed RNG with $\mathrm{hash}(s, v_1, v_2, \boldsymbol{x}_{-h:})$
    \STATE $r \gets$ RNG output from $(0,1)$
    \IF{$r <= 0.5$}
        \STATE $\mathcal{V}^g \gets \mathcal{V}^g \cup \{v_1\}$, \; $\mathcal{V}^r \gets \mathcal{V}^r \cup \{v_2\}$
    \ELSE
        \STATE $\mathcal{V}^g \gets \mathcal{V}^g \cup \{v_2\}$, \; $\mathcal{V}^r \gets \mathcal{V}^r \cup \{v_1\}$
    \ENDIF
\ENDFOR
\STATE $T \gets \gamma*|\mathcal{V}|-k/2$
\STATE Seed RNG with $\mathrm{hash}(s, \boldsymbol{x}_{-k:})$
\STATE $\mathcal{V}^* \gets$ Permutate $\mathcal{V}_{k:}$ seed by RNG
\STATE $\mathcal{V}^g \gets \mathcal{V}^g \cup \{\mathcal{V^*}_{0:T-1}\}$, \; 
\STATE $\mathcal{V}^r \gets \mathcal{V}^r \cup \{\mathcal{V^*}_{T:}\}$

\FOR{each $v_i \in \mathcal{V}^g$}
    \STATE $\tilde{L}_{i} \gets L_{i} + \delta$
\ENDFOR
\STATE $\tilde{P} \gets \mathrm{softmax}(\tilde{L})$
\STATE Sample next token $y \sim \tilde{P}$
\RETURN $y,\mathcal{V}^g,\mathcal{V}^r$
\end{algorithmic}
\end{algorithm}

\subsection{Watermark Detection}

As an algorithm for logit-balanced vocabulary partition, SSG can incorporate any KGW-scheme detection algorithm like \cite{kirchenbauer2024watermarklargelanguagemodels,lee2024wrotecodewatermarkingcode,lu2024entropybasedtextwatermarkingdetection,huang-etal-2025-low}.
In Algorithm \ref{alg:ssg_detection}, we adopt the design from \cite{lu2024entropybasedtextwatermarkingdetection} as an example for detection due to its reported high performance, which uses the entropy value of every token to determine its weight on final detection result.


\subsection{Analysis of SSG}
\label{AnalysisSSG}
In this section, we present an analysis of SSG to demonstrate why it can guarantee the lower bound of watermark strength even in low-entropy settings.

Let all $p_i$ be sorted descendingly: $p_1\ge p_2\ge p_3\ge\cdots \ge p_m \gg \sum_{i>m}p_i \approx 0$ and $p_{1:m}$ are high-probability tokens. SSG forms consecutive pairs $(p_1,p_2),(p_3,p_4),\ldots$ and assigns one token from each pair to $\mathcal{V}^g$.
Define
\begin{equation}
\alpha \;=\; \sum_{j} p_{2j}, 
\beta \;=\; \sum_{j} p_{2j-1}.    
\end{equation}

Then we have:
\begin{equation}
\alpha \;\ge\; \frac{1-p_{1}}{2}, 
\beta \;\le\; \frac{1+p_{1}}{2}.    
\end{equation}

Based on this, the lower bound $p_g^{lb}$ and upper bound $p_g^{ub}$ of $p_g$ are:
\begin{equation}
p_g^{lb}=\frac{1-p_{1}}{2} \le p_g \le \frac{1+p_{1}}{2}=p_g^{ub}.    
\end{equation}

Because $f_{ws}$ is a concave function about $p_g$, we then have: 
\begin{equation}
f_{ws}(p_g)\;\ge\min\{f_{ws}(p_g^{lb}),f_{ws}(p_g^{ub})\}>\;0.    
\end{equation}

which means $f_{ws}$ has a lower bound and it is determined by $p_1$. In this formulation, the value of $p_1$ directly controls the tightness of the lower bound. When $p_1$ is very large, the distribution is dominated by a single high-probability token. In this extreme case, the residual probability becomes small, resulting in a looser lower bound. Intuitively, if the model is already overwhelmingly confident in predicting one token, no watermarking scheme can substantially alter the distribution without comprising text quality, and thus the strength of the watermark signal is inherently limited.

Conversely, when $p_1$ is moderate, the probability is more evenly distributed among the top tokens. In this case, SSG ensures that both $\alpha$ and $\beta$ are bounded away from 0 and 1, leading to a tighter bound on $f_{ws}$. 

Taken together, this analysis demonstrates that SSG provides a provable guarantee of watermark strength that scales with the concentration of the next-token distribution. While extremely peaked distributions (large $p_1$) inherently limit the effectiveness of watermarking, SSG ensures that as long as probability is shared among multiple candidate tokens, the watermark signal remains detectable.

It is worth noting that, although SSG is designed to improve watermark detectability on low-entropy tasks, such analysis can still hold on high-entropy tasks. 
In the following, we validate its performance on both task types.

\begin{table*}[t]
\centering
\small
\begin{tabular}{lp{0.6cm}*{5}{>{\centering\arraybackslash}p{0.82cm}}|*{5}{>{\centering\arraybackslash}p{0.82cm}}}
\toprule
\multirow{2}{*}{\textbf{Model}} & \multirow{2}{*}{\textbf{Method}}
& \multicolumn{5}{c|}{\textbf{Humaneval}} 
& \multicolumn{5}{c}{\textbf{MBPP}} \\
& & P@1 & T@1 & F1@1 & T@5 & F1@5 & P@1 & T@1 & F1@1 & T@5 & F1@5 \\

\midrule
\multirow{7}{*}{Qwen2.5-coder-7B} 
 & \cellcolor{gray!25}UW & \cellcolor{gray!25}31.8 & \cellcolor{gray!25}0.6 & \cellcolor{gray!25}1.2 & \cellcolor{gray!25}3.7 & \cellcolor{gray!25}6.7 & \cellcolor{gray!25}42.3 & \cellcolor{gray!25}1.9 & \cellcolor{gray!25}3.6 & \cellcolor{gray!25}5.8 & \cellcolor{gray!25}10.5 \\
\cmidrule(l){2-12}
 & KGW         & 26.2 & 22.0 & 35.8 & 36.0 & 51.3 & 38.4 & 37.8 & 54.6 & 64.0 & 75.9 \\
 & $+\text{SSG}$ & 25.6 & 39.0  & 55.9 & 45.7 & 60.7  & 37.0  & 58.7  & 73.6  & 71.7  & 81.3  \\
\cmidrule(l){2-12}
 & SWEET       & 26.2 & 29.3 & 45.1 & 36.1 & 66.1 & 40.2 & 53.7 & 69.5 & 72.8 & 82.0 \\
 & $+\text{SSG}$ & 29.9 & 50.0 & 66.4 & \textbf{78.0} & \textbf{84.2} & 37.6 & 62.2 & 83.5 & 76.2 & 84.2 \\
\cmidrule(l){2-12}
 & EWD         & 26.5 & 32.9 & 49.3 & 52.4 & 66.7 & 37.6 & 57.4 & 75.7 & 74.6 & 83.2 \\
 & $+\text{SSG}$ & 27.8 & \textbf{50.6} & \textbf{66.9} & 70.7 & 80.6 & 38.1 & \textbf{74.3} & \textbf{84.9} & \textbf{84.1} & \textbf{89.1} \\
\midrule
\multirow{7}{*}{LLaMA-3-8B} 
 & \cellcolor{gray!25}UW & \cellcolor{gray!25}34.1 & \cellcolor{gray!25}8.5 & \cellcolor{gray!25}15.6 & \cellcolor{gray!25}15.2 & \cellcolor{gray!25}25.4 & \cellcolor{gray!25}56.3 & \cellcolor{gray!25}13.6 & \cellcolor{gray!25}18.3 & \cellcolor{gray!25}29.7 & \cellcolor{gray!25}10.5 \\
\cmidrule(l){2-12}
 & KGW         & 27.4 & 13.4 & 23.5 & 22.6 & 35.4 & 51.9 & 21.1 & 27.6 & 28.3 & 42.5 \\
 & $+\text{SSG}$ & 32.3 & 14.0 & 24.5 & 49.4 & 64.0 & 51.6 & 85.5 & 89.9 & 89.9 & 92.4 \\
\cmidrule(l){2-12}
 & SWEET       & 29.3 & 7.9 & 14.6 & 26.2 & 40.0 & 51.9 & 74.1 & 78.5 & 81.0 & 87.2 \\
 & $+\text{SSG}$ & 29.3 & 24.4 & 39.0 & 37.8 & 53.0 & 53.2 & 93.0 & 96.0 & 96.8 & 96.1 \\
\cmidrule(l){2-12}
 & EWD         & 34.1 & 26.8 & 42.1 & 54.9 & 68.7 & 51.6 & 87.7 & 91.3 & 94.2 & 94.7 \\
 & $+\text{SSG}$ & 29.9 & \textbf{59.1} & \textbf{74.0} & \textbf{73.2} & \textbf{82.2} & 51.6 & \textbf{97.1} & \textbf{98.1} & \textbf{98.9} & \textbf{97.1} \\
\bottomrule
\end{tabular}
\caption{Evaluation of watermarking methods on \textbf{Humaneval} and \textbf{MBPP}. For every setting, we run $5$ times and report the average result.
Numbers are percentages (i.e., original scores $\times 100$). \textbf{UW} denotes the unwatermarked baseline. 
\textbf{UW} denotes the unwatermarked baseline. 
Metrics are reported as percentages (original scores $\times 100$): 
\textbf{P@1} = Pass@1, 
\textbf{T@1} = True Positive Rate (TPR) at 1\% False Positive Rate (FPR), 
\textbf{F1@1} = F1-score at 1\% FPR, 
\textbf{T@5} = TPR at 5\% FPR, 
\textbf{F1@5} = F1-score at 5\% FPR. Best results in each column are bolded.}
\label{tab:code_results}
\end{table*}

\section{Experiments}

We conduct experiments across multiple datasets, models, and watermarking methods using the open-source toolkit MarkLLM \cite{pan2024markllmopensourcetoolkitllm}.  

\textbf{Tasks and Datasets.} We evaluate SSG on both low-entropy and high-entropy tasks. The two representative low-entropy tasks selected are code generation and mathematical problem solving. For code generation, we use HumanEval \cite{Humaneval} and MBPP \cite{MBPP}, both consisting of Python programming problems with reference solutions. For mathematical reasoning, we sample 10\% problems from GSM8K \cite{gsm8k}, which contains grade-school level math questions in English. For the high entropy tasks, we use C4 \cite{C4} and CNN/DailyMail \cite{cnn}. Following prior work \cite{lee2024wrotecodewatermarkingcode,lu2024entropybasedtextwatermarkingdetection,huang-etal-2025-low}, we only consider generations longer than 15 tokens for watermark detection. 

\textbf{Models.} For code generation, we employ Qwen2.5-Coder-7B\cite{hui2024qwen25codertechnicalreport}, a model specialized in programming tasks. For mathematical reasoning, we use DeepSeekMath-7B\cite{shao2024deepseekmathpushinglimitsmathematical}, which is trained specifically for math problem solving. To assess the generalizability of SSG, we also evaluate the versatile LLaMA-3-8B \cite{grattafiori2024llama3herdmodels} on both code and math tasks. The max generation lengyh is set to $512$ in all experiments.

\textbf{Watermarking Methods.} We compare three representative watermarking methods: KGW, SWEET, and EWD. KGW is the classical baseline in this line of research, while SWEET and EWD are methods designed to better handle low-entropy scenarios. Without lossing generality, we set watermark context window $h$ as $1$, $\delta$ as $2.0$, green set ratio $\gamma$ as $0.5$ for all methods. 

\textbf{Baselines and Metrics.} Since SSG can be incorporated into any KGW-scheme watermarking method, we report results for each method both with and without SSG, thereby isolating the effect of our contribution. Following prior works \cite{lu2024entropybasedtextwatermarkingdetection,lee2024wrotecodewatermarkingcode}, we report true positive rates (TPR) at 1\% and 5\% false positive rates (FPR), along with the corresponding F1-score. For generation quality, we measure Pass@1 accuracy on HumanEval and MBPP for code generation, and answer accuracy on GSM8K for mathematical reasoning. Besides, we use perplexity to measure text quality on C4 and CNN/DailyMail. These metrics together reflect the trade-off between watermark detectability and text quality.

\subsection{Main Results}

\begin{table}[t]
\centering
\small
\begin{tabular}{p{1.4cm} p{0.9cm} *{5}{>{\centering\arraybackslash}p{0.5cm}}}
\toprule
\textbf{Model} & \textbf{Method} 
& P@1 & T@1 & F1@1 & T@5 & F1@5 \\
\midrule
\multirow{7}{*}{DSMath-7B} 
 & \cellcolor{gray!25}UW & \cellcolor{gray!25}27.3 & \cellcolor{gray!25}4.5 & \cellcolor{gray!25}8.6 & \cellcolor{gray!25}6.8 & \cellcolor{gray!25}12.2 \\
\cmidrule(lr){2-7}
 & KGW         & 21.7 & 41.7 & 58.5 & 57.6 & 71.0 \\
 & $+\text{SSG}$ & 25.8 & 90.9 & 94.9 & 91.7 & 93.4 \\
\cmidrule(lr){2-7}
 & SWEET       & 27.3 & 94.7 & 96.9 & \textbf{97.0} & \textbf{96.2} \\
 & $+\text{SSG}$ & 25.0 & 94.7 & 96.9 & \textbf{97.0} & \textbf{96.2} \\
\cmidrule(lr){2-7}
 & EWD         & 21.2 & 93.2 & 96.1 & 95.5 & 95.5 \\
 & $+\text{SSG}$  & 25.0 & \textbf{96.2} & \textbf{97.7} & \textbf{97.0} & \textbf{96.2} \\
\midrule
\multirow{7}{*}{LLaMA-3-8B} 
 & \cellcolor{gray!25}UW & \cellcolor{gray!25}38.6 & \cellcolor{gray!25}5.3 & \cellcolor{gray!25}10.0 & \cellcolor{gray!25}12.9 & \cellcolor{gray!25}21.9 \\
\cmidrule(lr){2-7}
 & KGW         & 37.1 & 63.6 & 77.4 & 65.2 & 76.8 \\
 &$+\text{SSG}$  & 34.1 & 89.4 & 94.0 & 91.0 & 95.5 \\
\cmidrule(lr){2-7}
 & SWEET       & 38.6 & 79.5 & 88.2 & 94.7 & 95.1 \\
 & $+\text{SSG}$ & 34.1 & \textbf{100} & \textbf{99.6} & \textbf{100} & \textbf{97.8} \\
\cmidrule(lr){2-7}
 & EWD         & 35.6 & 74.2 & 84.8 & 95.5 & 95.5 \\
 & $+\text{SSG}$ & 35.6 & 99.2 & 99.2 & 99.2 & 97.4 \\
\bottomrule
\end{tabular}
\caption{Evaluation of watermarking methods on \textbf{GSM8K}. }
\label{tab:gsm8k_results}
\end{table}

\begin{figure*}[t]
  \centering
  \includegraphics[page=1,width=\textwidth]{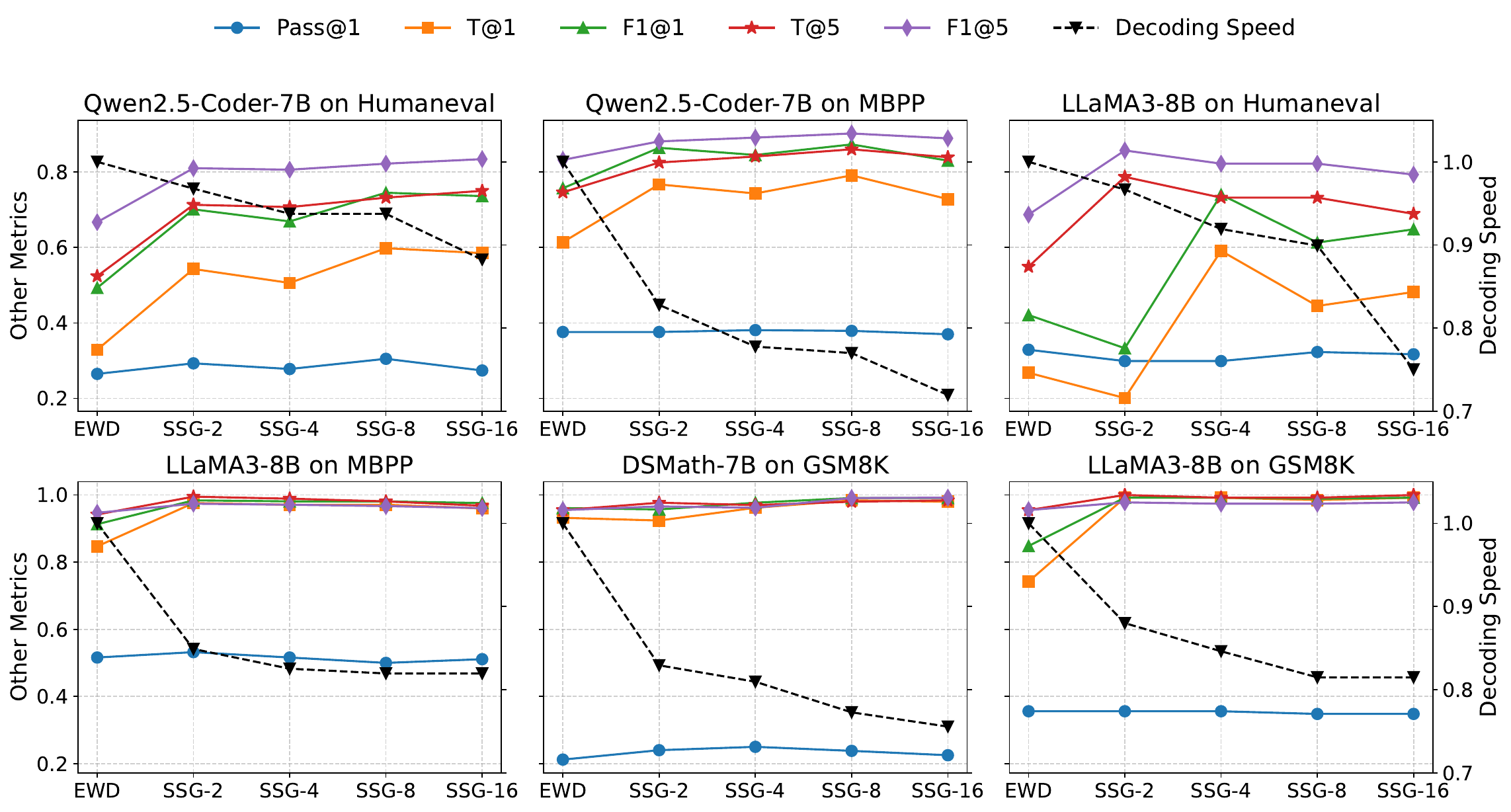}
  \caption{Influence of top-$k$ on SSG performance.}
  \label{fig:topkpdf}
\end{figure*}

Table \ref{tab:code_results} and Table \ref{tab:gsm8k_results} show the performance of multiple watermark algorithms on code generation and mathematical reasoning tasks, respectively, while Table \ref{tab:high_entropy_results} shows the results on high-entropy tasks. Across all models and datasets, the introduction of watermarking results in a drop in Pass@1 relative to the unwatermarked baseline. This indicates that watermarking inevitably perturbs the next-token distribution, and such perturbations come at a cost of text quality. In other words, watermarking strengthens attribution and traceability but reduces the probability of generating the correct solution.

For all three watermarking methods (KGW, SWEET, and EWD), the corresponding SSG variants yield substantial improvements in detection performance. In particular, SSG consistently raises TPR and F1 under both the $1\%$ and $5\%$ FPR thresholds, in some cases by large margins. Importantly, these gains are achieved without further degrading Pass@1, demonstrating that SSG can significantly strengthen watermark detectability while preserving text quality.

\begin{figure}[!htbp] 
    \centering
    \includegraphics[width=\columnwidth]{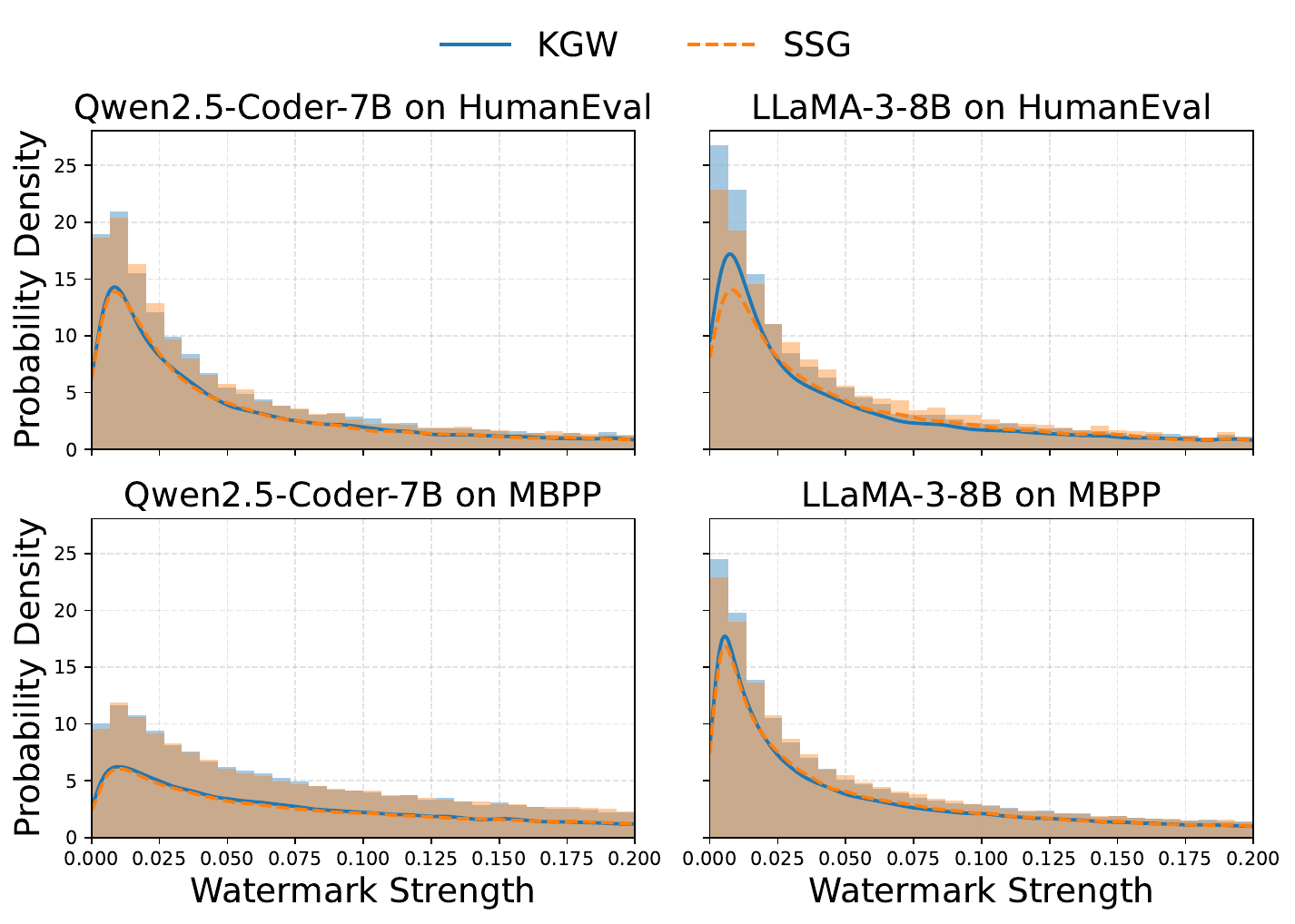}
    \caption{Examples of the watermark strength distributions on Qwen2.5-Coder-7B and LLaMA3-8B for code generation tasks.}
    \label{fig:ws}
\end{figure}
\subsection{Analysis on Watermark Strength}

To further assess the effect of SSG on token-level watermark detectability, we analyze the distribution of Watermark Strength on code generation tasks with Qwen2.5-Coder-7B and LLaMA-3-8B. Figure~\ref{fig:ws} illustrates the comparison between KGW and SSG. The key difference is that KGW produces a substantially larger proportion of tokens with near-zero Watermark Strength. Such tokens contribute little to the statistical signal and ultimately lower the overall $z$-score of the sequence, impairing detection. By contrast, SSG reshapes the distribution by reducing the probability of near-zero values and enforcing a lower bound on Watermark Strength, thereby ensuring more consistent detectability across tokens.

\begin{table*}[t]
\centering
\small
\begin{tabular}{llccccc|ccccc}
\toprule
\multirow{2}{*}{\textbf{Model}} & \multirow{2}{*}{\textbf{Method}}
& \multicolumn{5}{c|}{\textbf{Humaneval}} 
& \multicolumn{5}{c}{\textbf{MBPP}} \\
& & P@1 & T@1 & F1@1 & T@5 & F1@5 & P@1 & T@1 & F1@1 & T@5 & F1@5 \\

\midrule
\multirow{7}{*}{Qwen2.5-coder-7B} 
 & \cellcolor{gray!25}UW & \cellcolor{gray!25}31.8 & \cellcolor{gray!25}0.6 & \cellcolor{gray!25}1.2 & \cellcolor{gray!25}3.7 & \cellcolor{gray!25}6.7 & \cellcolor{gray!25}42.3 & \cellcolor{gray!25}1.9 & \cellcolor{gray!25}3.6 & \cellcolor{gray!25}5.8 & \cellcolor{gray!25}10.5 \\
\cmidrule(l){2-12}
 & KGW         & 26.2 & 22.0 & 35.8 & 36.0 & 51.3 & 38.4 & 37.8 & 54.6 & 64.0 & 75.9 \\
 & $+\text{SSG}$ & 25.6 & 22.6 & 36.6 & 31.7 & 46.4 & 37.0 & 55.6 & 71.1 & 67.5 & 78.3 \\
\cmidrule(l){2-12}
 & SWEET       & 26.2 & 32.3 & 48.6 & 53.0 & 67.2 & 40.2 & 57.1 & 72.4 & 72.2 & 81.6 \\
 & $+\text{SSG}$ & {29.9} & 30.5 & 46.5 & {54.3} & {68.2} & 37.6 & \textbf{72.2} & \textbf{80.2} & \textbf{75.7} & \textbf{83.7} \\
\cmidrule(l){2-12}
 & EWD         & 26.5 & \textbf{43.3} & \textbf{60.2} & \textbf{59.1} & \textbf{72.1} & 37.6 & 67.5 & \textbf{80.2} & 75.4 & \textbf{83.7} \\
 & $+\text{SSG}$ & 27.8 & 28.0 & 43.6 & 42.1 & 57.3 & {38.1} & {68.3} & {71.7} & {68.3} & {78.9} \\
\midrule
\multirow{7}{*}{LLaMA-3-8B} 
 & \cellcolor{gray!25}UW & \cellcolor{gray!25}32.9 & \cellcolor{gray!25}8.5 & \cellcolor{gray!25}15.6 & \cellcolor{gray!25}15.2 & \cellcolor{gray!25}25.4 & \cellcolor{gray!25}56.3 & \cellcolor{gray!25}13.6 & \cellcolor{gray!25}18.3 & \cellcolor{gray!25}29.7 & \cellcolor{gray!25}29.7 \\
\cmidrule(l){2-12}
 & KGW         & 27.4 & 13.4 & 23.5 & 22.6 & 35.4 & 51.9 & 22.1 & 27.6 & 28.3 & 42.5 \\
 & $+\text{SSG}$ & {32.3} & 6.7 & 12.5 & 23.2 & 36.2 & 51.6 & 60.8 & 80.0 & 82.5 & 88.1 \\
\cmidrule(l){2-12}
 & SWEET       & 29.3 & 8.5 & 15.6 & 18.3 & 29.7 & 51.9 & 41.8 & 58.6 & 60.1 & 72.9 \\
 & $+\text{SSG}$ & 29.3 & 10.4 & 18.7 & 14.6 & 24.5 & {53.2} & \textbf{94.5} & \textbf{94.5} & \textbf{91.3} & \textbf{93.1} \\
\cmidrule(l){2-12}
 & EWD         & {34.1} & 18.9 & 31.6 & \textbf{42.7} & \textbf{57.9} & 51.6 & 82.9 & 82.9 & 83.1 & 88.5 \\
 &$+\text{SSG}$ & 29.9 & \textbf{19.5} & \textbf{32.5} & 34.8 & 49.8 & 51.6 & 85.8 & 85.8 & 87.3 & 90.9 \\
\bottomrule
\end{tabular}
\caption{Evaluation of watermarking methods on \textbf{Humaneval} and \textbf{MBPP} without original prompts. All numbers are percentages; bold indicates the best per column.}
\label{tab:code_results_nogtp}
\end{table*}

\begin{table}[t]
\centering
\small
\begin{tabular}{p{1.4cm} p{0.9cm} *{5}{>{\centering\arraybackslash}p{0.5cm}}}
\toprule
\textbf{Model} & \textbf{Method} 
& P@1 & T@1 & F1@1 & T@5 & F1@5 \\
\midrule
\multirow{7}{*}{DSMath-7B} 
 & \cellcolor{gray!25}UW & \cellcolor{gray!25}27.3 & \cellcolor{gray!25}4.5 & \cellcolor{gray!25}8.6 & \cellcolor{gray!25}6.8 & \cellcolor{gray!25}12.2 \\
\cmidrule(lr){2-7}
 & KGW         & 21.7 & 41.7 & 58.5 & 57.6 & 71.0 \\
 & $+\text{SSG}$ & 25.8 & 81.8 & 89.6 & 85.6 & 90.0 \\
\cmidrule(lr){2-7}
 & SWEET       & \textbf{27.3} & \textbf{94.7} & \textbf{96.9} & \textbf{96.2} & \textbf{95.8} \\
 & $+\text{SSG}$ & 25.0 & 78.0 & 87.3 & 82.6 & 88.3 \\
\cmidrule(lr){2-7}
 & EWD         & 21.2 & 92.4 & 95.7 & 95.5 & 95.5 \\
 & $+\text{SSG}$  & 25.0 & 86.4 & 92.3 & 88.6 & 91.8 \\
\midrule
\multirow{7}{*}{LLaMA-3-8B} 
 & \cellcolor{gray!25}UW & \cellcolor{gray!25}35.6 & \cellcolor{gray!25}5.3 & \cellcolor{gray!25}10.0 & \cellcolor{gray!25}12.9 & \cellcolor{gray!25}21.9 \\
\cmidrule(lr){2-7}
 & KGW         & 35.6 & 63.6 & 77.4 & 65.2 & 76.8 \\
 & $+\text{SSG}$  & 34.1 & 71.2 & 82.8 & 78.8 & 86.0 \\
\cmidrule(lr){2-7}
 & SWEET       & \textbf{38.6} & 60.6 & 75.1 & 84.1 & 89.2 \\
 & $+\text{SSG}$ & 34.1 & \textbf{97.7} & \textbf{98.5} & \textbf{97.7} & \textbf{96.6} \\
\cmidrule(lr){2-7}
 & EWD         & 35.6 & 70.5 & 82.3 & 92.4 & 93.8 \\
 & $+\text{SSG}$ & 35.6 & 68.2 & 80.7 & 75.8 & 84.0 \\
\bottomrule
\end{tabular}
\caption{Evaluation of watermarking methods on \textbf{GSM8K}. All numbers are percentages; bold indicates the best per column.}
\label{tab:gsm8k_results_nogtp}
\end{table}

\subsection{The choice of Top-k}

We further investigate how the choice of top-$k$ in SSG influences watermarking performance, text quality, and decoding efficiency. 
Experiments are conducted across six settings: Qwen2.5-Coder-7B and LLaMA-3-8B on HumanEval and MBPP for code generation, and DSMath-7B and LLaMA-3-8B on GSM8K for mathematical reasoning. 
We compare the baseline EWD with its variants combined with SSG, varying $k \in \{2,4,8,16\}$ to control the number of top-logit tokens used for balanced vocabulary partitioning. 
For decoding efficiency, we normalize the speed of EWD to 100\%.  

The results in Figure~\ref{fig:topkpdf} show that even when SSG is applied only to the top-2 logits, it yields substantial improvements in detection performance (TPR@1, TPR@5, and F1) compared to the baseline. 
Increasing $k$ beyond 2 brings only marginal gains in detection while leaving text quality (Pass@1) unchanged, but incurs additional overhead that slows down decoding. 
Thus, larger $k$ values fail to provide further detectability benefits and instead reduce efficiency.  

Considering this trade-off, we recommend moderate settings of $k=2$ or $k=4$, which offer strong detection capability without significant efficiency loss.

\subsection{Results without Original Prompts}

In realistic scenarios such as code authorship detection, the original prompts used during generation may be unavailable at the time of watermark detection. To evaluate this setting, we follow prior work \cite{lu2024entropybasedtextwatermarkingdetection,lee2024wrotecodewatermarkingcode,huang-etal-2025-low} and conduct experiments using a general prompt. Table~\ref{tab:code_results_nogtp} and Table~\ref{tab:gsm8k_results_nogtp} report the detection results under this condition for code generation and mathematical reasoning, respectively. Compared to the case where original prompts are accessible, detection performance generally decreases across all methods, except for KGW without SSG. This is expected since KGW is the only method that does not depend on the original prompt for detection. Moreover, the effect of incorporating SSG is mixed: it improves detection in some cases but degrades it in others, reflecting the inherent reliance of SSG on logit values conditioned by the original prompts.

\subsection{Results on Robustness}

We also evaluate the robustness of SSG against \textbf{paraphrase attack} by using GPT-4o-mini to paraphrase watermarked sentences. The experiment results showed in Table \ref{tab:attack_results_code} and \ref{tab:attack_results_math} demonstrated that the robustness of SSG varies on different LLMs compared to other methods without SSG. Compared to the results in Table \ref{tab:code_results} and Table \ref{tab:gsm8k_results}, we can find that TPR drops significantly in all settings, demonstrating that paraphrasing attack is an effective way to erase watermark. SSG exhibits strong robustness on LLaMA-3-8B across both code generation and mathematical reasoning tasks. Conversely, it shows poor robustness on DSMath-7B. As for Qwen2.5-Coder-7B, its robustness remains largely on par with the baseline.

\section{Conclusion}

In this paper, we introduced a token-level metric, \textit{Watermark Strength}, to evaluate watermark detectability at the granularity of individual tokens. Using this metric, we revealed why the KGW scheme suffers in low-entropy settings such as code generation and mathematical reasoning. To address this limitation, we proposed \textbf{SSG} (Sort-then-Split by Groups), a logit-balanced partitioning algorithm that improves the lower bound on Watermark Strength of watermark injection. Extensive experiments on multiple large language models across code and mathematic reasoning tasks demonstrate that SSG achieves consistently stronger detection performance while maintaining text quality.  

\section*{Limitations}

Although SSG proves effective in our evaluations, the experimental scope remains limited. Future work should expand to additional low-entropy domains and diverse datasets to further validate its performance. Moreover, while SSG improves watermark injection reliability, it still depends on the availability of the original prompt for optimal detection, which restricts its applicability in realistic settings. An important research direction is to design prompt-free watermarking algorithms that retain effectiveness even in low-entropy tasks.



\bibliography{custom}

\clearpage
\appendix

\section{Watermark Detection}

Algorithm \ref{alg:ssg_detection} explains how SSG works with another detection algorithm like EWD.

\begin{algorithm}[t]
\caption{SSG Watermark Detection}
\label{alg:ssg_detection}
\begin{algorithmic}[1]
\REQUIRE LLM $\mathcal{M}$, suspected text $\boldsymbol{x}$, vocabulary $\mathcal{V}$, secret key $s$, threshold $\tau$, green-list ratio $\gamma$, context window size $h$, Top-k hyperparameter $k$
\ENSURE Decision on whether $\boldsymbol{x}$ is watermarked
\STATE Initialize $\boldsymbol{E} \gets \boldsymbol{0}$, $\boldsymbol{g} \gets \boldsymbol{0}$
\FOR{each token $x_i$ in $\boldsymbol{x}_{k:}$}
    \STATE $L \gets \mathcal{M}(\boldsymbol{x}_{0:i-1})$ \hfill 
    \STATE $P \gets \mathrm{softmax}(L)$
    \STATE $E_i \gets \text{Entropy}(P)$
    \STATE $(\mathcal{V}^g,\mathcal{V}^r) \gets \mathrm{SSG}(\mathcal{M}, \mathcal{V}, s, \boldsymbol{x}, h,k)$
    \IF{$x_i \in \mathcal{V}^g$}
        \STATE $\boldsymbol{g}_i \gets 1$
    \ENDIF
\ENDFOR
\FOR{each $E_i$ in $\boldsymbol{E}$}
    \STATE $W_i \gets E_i - \min(\boldsymbol{E})$
\ENDFOR
\STATE $|s|_G \gets \boldsymbol{W} \cdot \boldsymbol{g}$
\STATE $z \gets \dfrac{|s|_G - \gamma \sum_{i=h}^{|\boldsymbol{x}|} W_i}{\sqrt{\gamma(1-\gamma) \sum_{i=h}^{|\boldsymbol{x}|} W_i^2}}$
\IF{$z > \tau$}
    \RETURN $\boldsymbol{x}$ is watermarked
\ELSE
    \RETURN $\boldsymbol{x}$ is not watermarked
\ENDIF
\end{algorithmic}
\end{algorithm}

\section{Curve of Watermark Strength}
\label{gws}
\begin{figure}[!htbp] 
    \centering
    \includegraphics[width=\columnwidth]{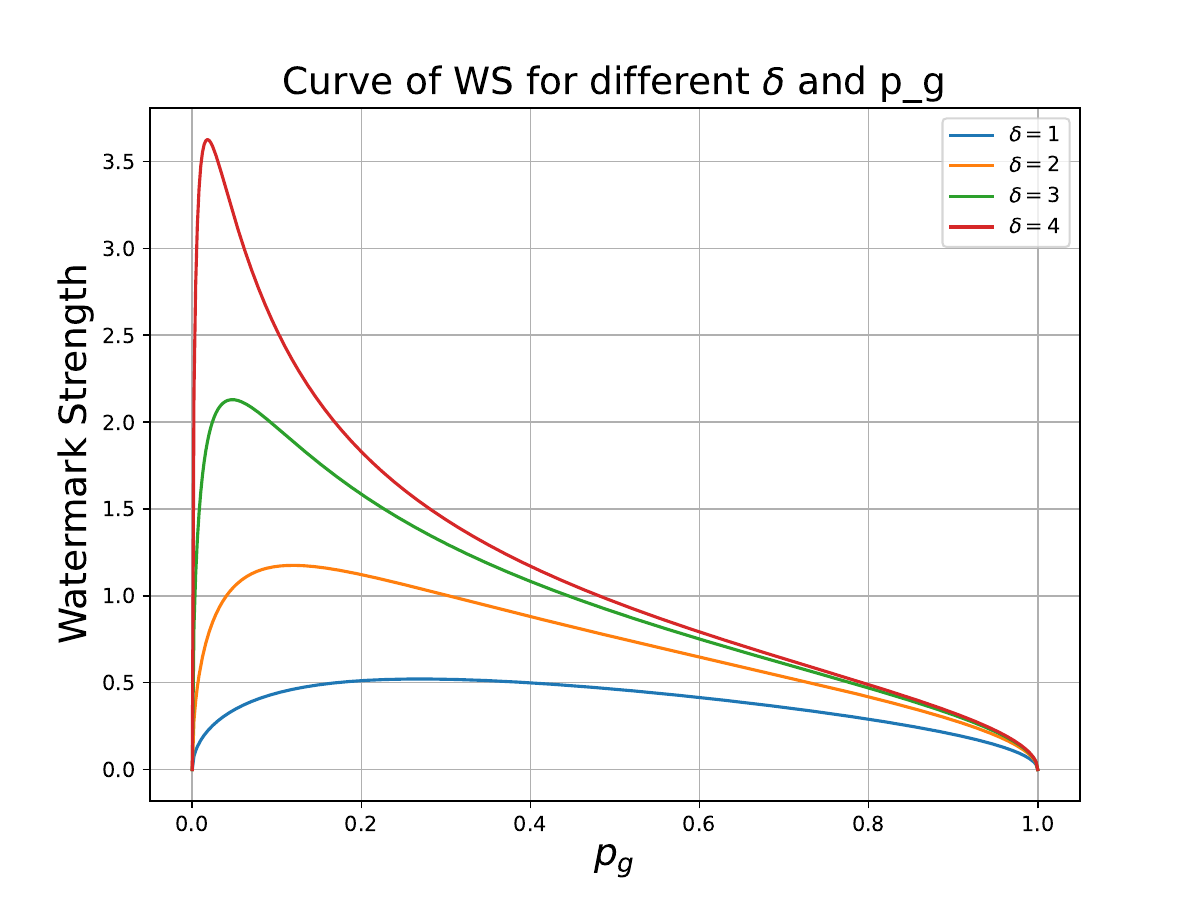}
    \caption{Curve of Watermark Strength over $p_g$ and $\delta$.}
    \label{fig:ws}
\end{figure}

\section{Text Quality of SSG}

We show that SSG and KGW are equivalent in the expected logit shift across the parameter space $\gamma \in (0,1)$. Let $\mathbb{E}[\tilde{L}_i(\gamma)]$ be the expected logit of token $v_i$ for a given $\gamma$.

In the KGW framework, the expectation is trivially:
\begin{equation}
    \mathbb{E}_{KGW}[\tilde{L}_i(\gamma)] = L_i + \gamma \delta
\end{equation}

In SSG, the probability of green tokens $P(v_i \in \mathcal{V}^g; \gamma)$ is piecewise-defined by the token's rank $Z_i$. The total bias injected into the vocabulary $\mathcal{V}$ is:
\begin{align}
    {SSG}(\gamma) &= \sum_{v_j \in \mathcal{V}} \left( \mathbb{E}_{SSG}[\tilde{L}_j(\gamma)] - L_j \right) \\
    &= \sum_{v_j \in \mathcal{V}_{top}} 0.5\delta + \sum_{v_j \in \mathcal{V}_{rem}} \frac{\gamma|\mathcal{V}| - k/2}{|\mathcal{V}|-k}\delta \\
    &= \frac{k}{2}\delta + (\gamma|\mathcal{V}| - k/2)\delta = \gamma|\mathcal{V}|\delta
\end{align}

Dividing by the vocabulary size $|\mathcal{V}|$, the mean expected logit for SSG is:
\begin{equation}
    \frac{1}{|\mathcal{V}|} \sum_{v_j \in \mathcal{V}} \mathbb{E}_{SSG}[\tilde{L}_j(\gamma)] = \bar{L} + \gamma \delta
\end{equation}

Thus, the global expectation of the logit shift in SSG is identical to KGW:
\begin{equation}
    \mathbb{E}_{SSG}[\tilde{L}(\gamma)] = \mathbb{E}_{KGW}[\tilde{L}(\gamma)] = L + \gamma \delta
\end{equation}
This confirms that SSG preserves the quality quality of KGW.

\section{General Prompts}
For Humaneval and MBPP, we use "Please write a Python function that solves a given task.\textbackslash n Output only valid Python code without explanations. \textbackslash n\textbackslash n def solve\_problem(input\_data): \textbackslash n" as our General Prompt for experiments without original prompts. While on GSM8k, the General Prompt is "Solve the following math word problem step by step, and in the last line output the final numeric answer in the format: \textbackslash n \#\#\#\# <answer>.\textbackslash n".

\section{Experiments}
\begin{table*}[t]
\centering
\small
\begin{tabular}{llccccc|ccccc}
\toprule
\multirow{2}{*}{\textbf{Model}} & \multirow{2}{*}{\textbf{Method}}
& \multicolumn{5}{c|}{\textbf{C4}} 
& \multicolumn{5}{c}{\textbf{CNN/DM}} \\
& & PPL & T@1 & F1@1 & T@5 & F1@5 & PPL & T@1 & F1@1 & T@5 & F1@5 \\

\multirow{7}{*}{LLaMA-3-8B} 
 & \cellcolor{gray!25}UW & \cellcolor{gray!25}2.65 & \cellcolor{gray!25}3.0 & \cellcolor{gray!25}5.8 & \cellcolor{gray!25}4.0 & \cellcolor{gray!25}7.3 & \cellcolor{gray!25}2.60 & \cellcolor{gray!25}0.0 & \cellcolor{gray!25}0.0 & \cellcolor{gray!25}2.0 & \cellcolor{gray!25}3.7 \\
\cmidrule(l){2-12}
 & KGW         & 3.01 & 81.0 & 89.0 & 90.0 & 92.3 & 2.86 & 72.0 & 83.2 & 91.0 & 92.9 \\
 & $+\text{SSG}$ & {2.80} & 93.0 & 95.9 & 95.0 & 95.0 & 2.84 & 98.0 & 98.5 & 99.0 & 97.1 \\
\cmidrule(l){2-12}
 & SWEET       & 3.01 & 91.0 & 94.8 & 93.0 & 93.9 & 2.89 & 88.0 & 93.1 & 99.0 & 97.1 \\
 & $+\text{SSG}$ & 2.80 & 94.0 & 96.8 & 95.2 & 95.8 & {2.84} & 95.0 & 95.0 & 99.0 & 97.1 \\
\cmidrule(l){2-12}
 & EWD         & {3.01} & 95.0 & 96.9 & 97.0 & 96.0 & 2.86 & 92.0 & 95.3 & 95.0 & 95.0 \\
 &$+\text{SSG}$ & 2.80 & 96.0 & 97.5 & 98.0 & 96.6 & 2.84 & 98.0 & 98.5 & 99.0 & 97.1 \\
\bottomrule
\end{tabular}
\caption{Evaluation of watermarking methods on \textbf{C4} and \textbf{CNN/DailyMail} . All numbers are percentages.}
\label{tab:high_entropy_results}
\end{table*}

\begin{table*}[t]
\centering
\small
\begin{tabular}{llcccc|cccc}
\toprule
\multirow{2}{*}{\textbf{Model}} & \multirow{2}{*}{\textbf{Method}}
& \multicolumn{4}{c|}{\textbf{Humaneval}} 
& \multicolumn{4}{c}{\textbf{MBPP}} \\
&  & T@1 & F1@1 & T@5 & F1@5  & T@1 & F1@1 & T@5 & F1@5 \\

\midrule
\multirow{7}{*}{Qwen2.5-coder-7B} 
 & \cellcolor{gray!25}UW & \cellcolor{gray!25}0.6 & \cellcolor{gray!25}1.2 & \cellcolor{gray!25}3.7 & \cellcolor{gray!25}6.7 & \cellcolor{gray!25}1.9 & \cellcolor{gray!25}3.6 & \cellcolor{gray!25}5.8 & \cellcolor{gray!25}10.5\\
\cmidrule(l){2-10}
 & KGW         & 3.0 & 5.9 & 9.1 & 16.1 & 3.2 & 6.1 & 14.3 & 24.0\\
\cmidrule(l){2-10}
 & SWEET       & 1.2 & 2.4 & 8.5 & 15.1 & 2.6 & 5.1 & 9.3 & 16.2 \\
\cmidrule(l){2-10}
 & EWD         & 5.5 & 10.3 & 11.0 & 18.9 & 4.8 & 9.0 & 8.7 & 15.4  \\
 & $+\text{SSG}$ &7.3 & 13.6 & 20.1 & 32.2 & 2.4 & {4.6} & {6.1} & {11.0} \\
\midrule
\multirow{7}{*}{LLaMA-3-8B} 
 & \cellcolor{gray!25}UW & \cellcolor{gray!25}6.1 & \cellcolor{gray!25}11.4 & \cellcolor{gray!25}11.6 & \cellcolor{gray!25}19.9 & \cellcolor{gray!25}1.6 & \cellcolor{gray!25}3.1 & \cellcolor{gray!25}4.5 & \cellcolor{gray!25}8.2  \\
\cmidrule(l){2-10}
 & KGW         & 0.6 & 1.2 & 1.2 & 2.3 & 1.1 & 2.1 & 5.0 & 9.2\\
\cmidrule(l){2-10}
 & SWEET       & 0.6 & 1.2 & 1.2 & 2.3 & 1.1 & 2.1 & 5.0 & 9.2 \\
\cmidrule(l){2-10}
 & EWD         & {7.3} & 13.6 & 23.8 & 37.0 & 4.0 & 7.6 & 12.7 & 21.6  \\
 &$+\text{SSG}$ & 28.0 & 43.6 & 53.0 & 67.2 & 5.6 & 10.4 & 19.6 & 31.5 \\
\bottomrule
\end{tabular}
\caption{Evaluation of watermarking methods on \textbf{Humaneval} and \textbf{MBPP} when watermarked sentence has been paraphrased by GPT-4o-mini. All numbers are percentages.}
\label{tab:attack_results_code}
\end{table*}

\begin{table}[t]
\centering
\small
\begin{tabular}{p{1.4cm} p{0.9cm} *{4}{>{\centering\arraybackslash}p{0.5cm}}}
\toprule
\textbf{Model} & \textbf{Method} 
 & T@1 & F1@1 & T@5 & F1@5 \\
\midrule
\multirow{5}{*}{DSMath-7B} 
 & \cellcolor{gray!25}UW & \cellcolor{gray!25}4.5 & \cellcolor{gray!25}8.6 & \cellcolor{gray!25}6.8 & \cellcolor{gray!25}12.2\\
\cmidrule(lr){2-6}
 & KGW         & 1.5 & 3.0 & 6.1 & 11.0  \\
\cmidrule(lr){2-6}
 & SWEET       & 25.0 & 39.8 & 44.7 & 59.9  \\
\cmidrule(lr){2-6}
 & EWD         & 22.7 & 36.8 & 28.8 & 43.2  \\
 & $+\text{SSG}$  & 0.0 & 0.0 & 9.8 & 17.2  \\
\midrule
\multirow{5}{*}{LLaMA-3-8B} 
 & \cellcolor{gray!25}UW & \cellcolor{gray!25}2.3 & \cellcolor{gray!25}4.4 & \cellcolor{gray!25}3.8 & \cellcolor{gray!25}7.0\\
\cmidrule(lr){2-6}
 & KGW         & 0.0 & 0.0 & 13.6 & 23.1  \\
\cmidrule(lr){2-6}
 & SWEET       & 0.0 & 0.0 & 13.6 & 23.1  \\
\cmidrule(lr){2-6}
 & EWD         & 1.5 & 3.0 & 12.1 & 20.8  \\
 & $+\text{SSG}$  & 12.1 & 21.5 & 25.0 & 38.6  \\
\bottomrule
\end{tabular}
\caption{Evaluation of watermarking methods on \textbf{GSM8K} when watermarked sentence has been paraphrased by GPT-4o-mini. All numbers are percentages.}
\label{tab:attack_results_math}
\end{table}

\end{document}